\documentstyle[editedvolume,numreferences]{crckapb}

\begin{opening}
\title{TIME, CLOSED TIMELIKE CURVES AND CAUSALITY}

\author{F. LOBO$^1$}

\author{P. CRAWFORD$^2$}

\institute{Centro de Astronomia e Astrof\'{\i}sica da Universidade de
Lisboa\\
           Campo Grande, Ed. C8 1749-016 Lisboa, Portugal}

\end{opening}

\runningtitle{TIME, CLOSED TIMELIKE CURVES AND CAUSALITY}

\begin{document}

\section{Introduction}

It seems to be extremely difficult to give a precise definition of
Time, this mysterious ingredient of the Universe. Intuitively, we
have the notion of time as something that flows. Ancient religions
have registered it as something unusual, and many myths are built
into their dogmas.

The ancient Greeks conveyed the image of Chronos, or Father Time.
Plato assumed that time had a beginning, looping back into itself.
This notion of circular time, was probably inspired by phenomena
observed in Nature, namely the alternation of day and night, the
repetition of the seasons, etc. But, it was in the Christian
theological doctrine that the unique character of historical
events gave rise to a linear notion of time. Aristotle, a keen
natural philosopher, stated that time was related to motion, i.e.,
to change. An idea reflected in his famous metaphor: {\it Time is
the moving image of Eternity}.

Reflections on time can be encountered in many philosophical
considerations and works over the ages, culminating in Newton's
Absolute Time. Newton stated that time flowed at the same rate for
all observers in the Universe. But in 1905, Einstein changed
altogether our notion of time. Time flowed at different rates for
different observers, and Minkowski, three years later, formally
united the parameters of time and space, giving rise to the notion
of a four-dimensional entity, spacetime.

Later, Einstein influenced by Mach's Principle, was motivated to
seek a theory in which the structure of spacetime was influenced
by the presence of matter, and presented the field equations of
the General Theory of Relativity in $1915$. Adopting a pragmatic
point of view, to measure time a changing configuration of matter
is needed, i.e., a swinging pendulum, etc. Change seems to be
imperative to have an emergent notion of time.

Therefore, time is empirically related to change. But change can
be considered as a variation or sequence of occurrences. Thus,
intuitively, a sequence of successive occurrences, provides us
with a notion of something that flows, i.e., it provides us with
the notion of {\it Time}. Time flows and everything relentlessly
moves along this stream.

In Relativity, we can substitute the above empirical notion of a
sequence of occurrences by a sequence of {\it events}. We idealize
the concept of an event to become a point in space and an instant
in time.

Following this reasoning, a sequence of events has a determined
{\it temporal order}. We experimentally verify that specific
events occur before others and not vice-versa. Certain events
(effects) are triggered off by others (causes), providing us with
the notion of {\it causality}.

\section{Closed Timelike Curves and Associated Paradoxes of Time Travel}

The conceptual definition and understanding of Time, both
quantitatively and qualitatively is of the utmost difficulty and
importance. Special Relativity provides us with important
quantitative elucidations of the fundamental processes related to
time dilation effects. The General Theory of Relativity (GTR)
provides a deep analysis to effects of time flow in the presence
of strong and weak gravitational fields.

As time is incorporated into the proper structure of the fabric of
spacetime, it is interesting to note that GTR is contaminated with
non-trivial geometries which generate {\it closed timelike curves}
\cite{Visser}. A closed timelike curve (CTC) allows time travel,
in the sense that an observer which travels on a trajectory in
spacetime along this curve, returns to an event which coincides
with the departure. The arrow of time leads forward, as measured
locally by the observer, but globally he/she may return to an
event in the past. This fact apparently violates causality,
opening Pandora's box and producing time travel paradoxes
\cite{Nahin}, throwing a veil over our understanding of the
fundamental nature of Time. The notion of causality is fundamental
in the construction of physical theories, therefore time travel
and its associated paradoxes have to be treated with great
caution. The paradoxes fall into two broad groups, namely the {\it
consistency paradoxes} and the {\it causal loops}.

The consistency paradoxes include the classical grandfather
paradox. Imagine travelling into the past and meeting one's
grandfather. Nurturing homicidal tendencies, the time traveller
murders his grandfather, preventing the birth of his father,
therefore making his own birth impossible. In fact, there are many
versions of the grandfather paradox, limited only by one's
imagination. The consistency paradoxes occur whenever
possibilities of changing events in the past arise.

The paradoxes associated with causal loops are related to
self-existing information or objects, trapped in spacetime.
Imagine a time traveller going back to his past, handing his
younger self a manual for the construction of a time machine. The
younger version then constructs the time machine over the years,
and eventually goes back to the past to give the manual to his
younger self. The time machine exists in the future because it was
constructed in the past by the younger version of the time
traveller. The construction of the time machine was possible
because the manual was received from the future. Both parts
considered by themselves are consistent, and the paradox appears
when considered as a whole. One might inquire as to the origin of
the manual, since its worldline is a closed loop. There is a
manual never created, nevertheless existing in spacetime, although
there are no causality violations.

\section{Solutions of the EFEs Generating CTCs}

A great variety of solutions to the Einstein Field Equations
(EFEs) containing CTCs exist, but two particularly notorious
features seem to stand out. Solutions with a tipping over of the
light cones due to a rotation about a cylindrically symmetric
axis; and solutions that violate the Energy Conditions of GTR,
which are fundamental in the singularity theorems and theorems of
classical black hole thermodynamics \cite{Hawking}.

\subsection{Stationary, axisymmetric solutions}

The tipping over of light cones seem to be a generic feature of
some solutions with a rotating cylindrical symmetry. The general
metric for a stationary, axisymmetric solution with rotation is
given by \cite{Visser,Wald}:
\begin{equation}
ds^2=-A(r)dt^2+2B(r)d\phi \,dt+C(r)d\phi ^2+D(r)(dr^2+dz^2)
\end{equation}
The range of the coordinates is: $t\in (-\infty, +\infty)$, $r\in
(0, +\infty)$, $\phi \in [0,2\pi]$, and $z \in (-\infty,
+\infty)$, respectively. The metric components are functions of
$r$ alone. It is clear that the determinant,
$g=det(g_{\mu\nu})=-(AC+B^2)D^2$ is Lorentzian, provided that
$(AC+B^2)>0$.

Due to the periodic nature of the angular coordinate, $\phi$, an
azimuthal curve with $\gamma =\{t={\it const},r={\it const},z={\it
const}\}$ is a closed curve of invariant length $s_{\gamma}^2
\equiv C(r)(2\pi)^2$. If $C(r)$ is negative then the integral
curve with $(t,r,z)$ fixed is a CTC.

The present work is far from making an exhaustive search of all
the EFE solutions generating CTCs with these features, but the
best known spacetimes will be briefly analyzed, namely, the van
Stockum spacetime, the G\"{o}del universe and spinning cosmic
strings.

\subsubsection{Van Stockum Spacetime}

The earliest solution to the EFEs containing CTCs, is probably
that of the van Stockum spacetime \cite{Visser,Tipler}. It is a
stationary, cylindrically symmetric solution describing a rapidly
rotating, infinitely long cylinder of dust, surrounded by vacuum.
The centrifugal forces of the dust are balanced by the
gravitational attraction. Consider $R$ the surface of the
cylinder.

The metric for the interior solution $r<R$, is given by:
\begin{equation}
ds^2=-dt^2+2\omega r^2 d\phi \, dt+r^2(1-\omega^2
r^2)d\phi^2+\exp(-\omega^2 r^2)(dr^2+dz^2)
\end{equation}
where $\omega$ is the angular velocity of the cylinder. It is
trivial to verify that CTCs arise if $\omega r>1$. Causality
violation can also be verified for $\omega R > 1/2$, in the
exterior region.

\subsubsection{Spinning Cosmic String}

Consider an infinitely long straight string that lies along and
spins around the $z$-axis. The symmetries are analogous to the van
Stockum spacetime, but the asymptotic behavior is different
\cite{Visser}.

We restrict the analysis to an infinitely long straight string,
with a delta-function source confined to the $z$-axis. It is
characterized by a mass per unit length, $\mu$; a tension, $\tau$,
and an angular momentum per unit length, $J$.

In cylindrical coordinates the metric takes the following form:
\begin{equation}
ds^2=-\left [d(t+4GJ\phi) \right ]^2+dr^2+(1-4G\mu)^2 \, r^2 \,
d\phi^2+dz^2
\end{equation}

Consider an azimuthal curve, i.e., an integral curve of $\phi$.
Closed timelike curves appear whenever $r<4GJ/(1-4G\mu)$.

\subsubsection{The G\"{o}del Universe}

Kurt G\"{o}del in $1949$ discovered an exact solution to the EFEs
of a uniformly rotating universe containing dust and a nonzero
cosmological constant. Writing the metric in a form in which the
rotational symmetry of the solution, around the axis $r=0$, is
manifest and suppressing the irrelevant $z$ coordinate, we have
\cite{Hawking,Godel}:
\begin{equation}
ds^2=2w^{-2} (-dt'^2+dr^2-(\sinh ^4r-\sinh ^2r)\,d\phi
^2+2(\sqrt{2})\sinh ^2r \,d\phi \,dt)
\end{equation}

Moving away from the axis, the light cones open out and tilt in
the $\phi$-direction. The azimuthal curves with $\gamma =\{t={\it
const},r={\it const},z={\it const}\}$ are CTCs if the condition
$r>\ln (1+\sqrt{2})$ is satisfied.

\subsection{Solutions violating the Energy Conditions}

The traditional manner of solving the EFEs, $G_{\mu \nu}=8\pi G
T_{\mu \nu}$, consists in considering a plausible stress-energy
tensor, $T_{\mu \nu}$, and finding the geometrical structure,
$G_{\mu\nu}$. But one can run the EFE in the reverse direction by
imposing an exotic metric $g_{\mu\nu}$, and eventually finding the
matter source for the respective geometry.

In this fashion, solutions violating the energy conditions have
been obtained. One of the simplest energy conditions is the weak
energy condition (WEC), which states:
$T_{\mu\nu}U^{\mu}U^{\nu}\geq 0$, in which $U^{\mu}$ is a timelike
vector. This condition is equivalent to the assumption that any
timelike observer measures a local positive energy density.
Although classical forms of matter obey these energy conditions,
violations have been encountered in quantum field theory, the
Casimir effect being a well-known example.

Adopting the reverse philosophy, solutions such as traversable
wormholes, the warp drive, the Krasnikov tube and the Ori-Soen
spacetime have been obtained. These solutions violate the energy
conditions and with simple manipulations generate CTCs.

\subsubsection{Traversable Wormholes, the Gott Spacetime and the
Ori-Soen Solution}

Much interest in traversable wormholes had been aroused since the
classical article by Morris and Thorne \cite{Morris}. A wormhole
is a hypothetical tunnel which connects different regions in
spacetime. These solutions are multiply-connected and probably
involve a topology change, which by itself is a problematic issue.
One of the most fascinating aspects of wormholes is their apparent
ease in generating CTCs. There are several ways to generate a time
machine using multiple wormholes \cite{Visser}, but a manipulation
of a single wormhole seems to be the simplest way \cite{MT}.

An extremely elegant model of a time-machine was constructed by
Gott \cite{Gott}. It is an exact solution to the EFE for the
general case of two moving straight cosmic strings that do not
intersect. This solution produces CTCs even though they do not
violate the WEC, have no singularities and event horizons, and are
not topologically multiply-connected as the wormhole solution. The
appearance of CTCs relies solely on the gravitational lens effect
and the relativity of simultaneity.

A time-machine model was also proposed by Amos Ori and Yoav Soen
which significantly ameliorates the conditions of the EFE's
solutions which generate CTCs \cite{Soen}. The Ori-Soen model
presents some notable features. It was verified that CTCs evolve
from a well-defined initial slice, a partial Cauchy surface, which
does not display causality violation. The partial Cauchy surface
and spacetime are asymptotically flat, contrary to the Gott
spacetime, and topologically trivial, contrary to the wormhole
solutions. The causality violation region is constrained within a
bounded region of space, and not at infinity as in the Gott
solution. The WEC is satisfied until and beyond a time slice
$t=1/a$, on which the CTCs appear.

\subsubsection{The Alcubierre Warp Drive and the Krasnikov Solution}

Within the framework of general relativity, it is possible to warp
spacetime in a small {\it bubblelike} region \cite{Alcubierre}, in
such a way that the bubble may attain arbitrarily large
velocities, $v(t)$. Inspired in the inflationary phase of the
early Universe, the enormous speed of separation arises from the
expansion of spacetime itself. The model for hyperfast travel is
to create a local distortion of spacetime, producing an expansion
behind the bubble, and an opposite contraction ahead of it.

One may consider a spaceship immersed within the bubble, moving
along a timelike curve, regardless of the value of $v(t)$. Due to
the arbitrary value of the warp bubble velocity, the metric of the
warp drive permits superluminal travel. This possibility raises
the question of the existence of CTCs. Although the solution
deduced by Alcubierre by itself does not possess CTCs, Everett
demonstrated that these are created by a simple modification of
the Alcubierre metric \cite{Everett}, by applying a similar
analysis as in tachyons.

Krasnikov discovered an interesting feature of the warp drive, in
which an observer in the center of the bubble is causally
separated from the front edge of the bubble. Therefore he/she
cannot control the Alcubierre bubble on demand. Krasnikov proposed
a two-dimensional metric \cite{Krasnikov}, which was later
extended to a four-dimensional model \cite{ER}. Using two such
tubes it is a simple matter, in principle, to generate CTCs.

\section{Conclusion}

GTR has been an extremely successful theory, with a well
established experimental footing, at least for weak gravitational
fields. Its predictions range from the existence of black holes
and gravitational radiation to the cosmological models, which
predict a primordial beginning, namely the big-bang.

However, it was seen that it is possible to find solutions to the
EFEs, with certain ease, which generate CTCs. This implies that if
we consider GTR valid, we need to include the {\it possibility} of
time travel in the form of CTCs. A typical reaction is to exclude
time travel due to the associated paradoxes. But the paradoxes do
not prove that time travel is mathematically or physically
impossible. Consistent mathematical solutions to the EFEs have
been found, based on plausible physical processes. What they do
seem to indicate is that local information in spacetimes
containing CTCs are restricted in unfamiliar ways.

The grandfather paradox, without doubt, does indicate some strange
aspects of spacetimes that contain CTCs. It is logically
inconsistent that the time traveller murders his grandfather. But,
one can ask, what exactly prevented him from accomplishing his
murderous act if he had ample opportunities and the free will to
do so. It seems that certain conditions in local events are to be
fulfilled for the solution to be globally self-consistent. These
conditions are denoted {\it consistency constraints}
\cite{Earman}. To eliminate the problem of free will, mechanical
systems were developed, such as the self-collision of billiard
balls in the presence of CTCs \cite{Echeverria}. These do not
convey the associated philosophical speculations on free will
related to human beings. Much has been written on two possible
remedies to the paradoxes, namely the Principle of
Self-Consistency and the Chronology Protection Conjecture.

Igor Novikov is a leading advocate for the Principle of
Self-Consistency, which stipulates that events on a CTC are
self-consistent, i.e., events influence one another along the
curve in a cyclic and self-consistent way. In the presence of CTCs
the distinction between past and future events is ambiguous, and
the definitions considered in the causal structure of well-behaved
spacetimes break down. What is important to note is that events in
the future can influence, but cannot change, events in the past.

The Principle of Self-Consistency permits one to construct local
solutions of the laws of physics, only if these can be prolonged
to a unique global solution, defined throughout non-singular
regions of spacetime. Therefore, according to this principle, the
only solutions of the laws of physics that are allowed locally,
reinforced by the consistency constraints, are those which are
globally self-consistent.

Hawking's Chronology Protection Conjecture is a more conservative
way of dealing with the paradoxes. Hawking notes the strong
experimental evidence in favour of the conjecture from the fact
that "we have not been invaded by hordes of tourists from the
future" \cite{Hawk2}.

An analysis reveals that the renormalized expectation value of the
quantum stress-energy tensor diverges as one gets close to CTC
formation. This conjecture permits the existence of traversable
wormoles, but prohibits the appearance of CTCs. The transformation
of a wormhole into a time machine results in enormous effects of
the vacuum polarization, which destroys its internal structure.
There is no convincing demonstration of the Chronology Protection
Conjecture, but the hope exists that a future theory of quantum
gravity may prohibit CTCs.

In addition to these remedies, Visser considers two other
conjectures \cite{Visser}. The first is the {\it radical
reformulation of physics conjecture}, in which one abandons the
causal structure of the laws of physics and allows, without
restriction, time travel, reformulating physics from the ground
up. The second is the {\it boring physics conjecture}, in which
one simply ceases to consider the solutions to the EFEs generating
CTCs.

Perhaps an eventual quantum gravity theory will provide us with
the answers. But, as stated by Thorne \cite{Thorne}, it is by
extending the theory to its extreme predictions that one can get
important insights to its limitations, and probably ways to
overcome them. Therefore, time travel in the form of CTCs is more
than a justification for theoretical speculation, it is a
conceptual tool and an epistemological instrument to probe the
deepest levels of GTR and extract clarifying views.

\end{document}